# Artificial Intelligence, connected products, virtual reality: potential impacts on user/consumer safety in terms of their physical and psychological ability or well-being

Birgitta Dresp-Langley

*Centre National de la Recherche Scientifique CNRS, France*

**Abstract**

With the progressive digitalisation of a majority of services to communities and individuals, humankind is facing new challenges. While energy sources are rapidly dwindling and rigorous choices have to be made to ensure the sustainability of our environment, there is increasing concern in science and society about the safety of connected products and technology for the individual user. This essay provides a first basis for further inquiry into the risks in terms of potentially negative, short and long-term, effects of connected technologies and massive digitalisation on the psychological and/or physical abilities and well-being of users or consumers.

**Introduction**

With what von der Malsburg **[1]** called 'the rise of Artificial Intelligence (AI)' and the progressive digitalisation of a majority of services to communities and individuals, humankind is facing new challenges that need to be met. First, AI and massive digitalisation have increased, and will further increase, our carbon foot print by orders of magnitude **[2]** that are largely unreasonable in the light of the fact that energy sources are rapidly dwindling. Rigorous choices will have to be made here to ensure the sustainability of our environment. Moreover, there is increasing concern in science and society about the *safety* of some of this new technology for the individual user, and for society as a whole.

This leads us directly to the question: *what is a safe product?* As a working definition, we herein suggest that **a safe product will not negatively affect either the privacy, or the physical, cognitive, and/or psychological ability or well-being of an individual who uses the product actively, or is exposed to it passively.** Any product that does not fulfil all of these criteria represents a risk to the user/consumer.

With this definition in mind, the following text provides a brief and comprehensive, albeit certainly not (yet) complete, overview of this complex problem space. As starting point, the author proposes a list of results from recent research efforts. No specific order, of priority or other, of the items listed is to be inferred. This preliminary summary points towards several, clearly identified, negative effects of digital (as synonymous with connected) technology applications and provides a first basis for further inquiries into the risks of connected technology in terms of potentially negative short and long-term effects on the psychological and/or physical well-being of users/consumers.

> 1- *How do you measure up today? Digital feed-back systems and the negative psychological effects of user performance rankings*

This point relates directly to social media in general and also to more specific online user exchange platforms where direct feed-back is given to the user/consumer by the computer generating some kind of assessment of how the user "measures up" in the system. Some individual consumers may draw psychological benefits from such competition-oriented direct feed-back and the online "ranking" of their contributions while others will be put off significantly by such feed-back, may even perceive it as a form of "harassment", become stressed, depressed, or even suicidal as a consequence of diminished self-esteem. A recent study published in the American Journal of Epidemiology has shown that teenagers are particularly vulnerable in this respect, and that the trend towards technology-induced teenage depression and associated symptoms may well reach epidemic proportions if nothing gets done to stop it **[3]**. Quite clearly, the manner in which individuals are led to interact with a computer controlled system in a given context can have negative short and long-term effects on their well-being and overall behaviour, including online behaviours. It is suggested that some types of online feed-back can undermine users'/consumers' positive attitude towards a given product, service provider, or group. This can then alter their motivation to behave positively and constructively **[4]**, within the computer controlled system, and within society as a whole. The risk of episodic or sustained psychological stress caused by inadequate feed-back systems, producing symptoms of depression and/or inappropriate and/or dangerous behaviours in individuals can easily be mitigated. Regulatory measures constraining companies/developers to test their interactive systems in this regard before they are implemented will be useful here and such tests should be performed on sufficiently large consumer populations. Since so many interactive systems are already out there, user testing and feed-back restrictions should be imposed for these.

## 2- *Connected technologies and the negative effects of multitasking on cognitive ability*

Apart from the fact that some recent work points towards a significant link between multitasking and a risk for obesity in young adults **[5]**, there is even more concerning evidence that multitasking is linked to poorer performance in a number of cognitive ability tests **[6].** The simultaneous processing of multiple incoming streams of information is a challenge for the human cognitive system, which has limited short-term memory capacity. To commit information to memory successfully, we need to be able to process information selectively; to ensure that only relevant information gets committed to memory, we need the cognitive ability to distinguish what is relevant from what is not in a multitude of incoming information. Heavy multitasking appears to make us more susceptible to interference from irrelevant environmental stimuli and irrelevant representations in memory and, therefore, we may lose our ability to correctly select relevant input when necessary. This cognitive deficit has previously been termed "not being able to see the trees for the forest" by neuroscientists **[7]**. Thus, all the multitasking that inevitably goes along with connected technologies incurs the risk that we may lose our selective information processing capability and, as a consequence, our critical ability for priority-based decision making. Cleary, processing whole streams of mostly irrelevant information online is unnecessary and does not make good use of our precious time. It appears therefore mandatory to inform users and consumers of the risks of multitasking to their selective information processing capacities and, ultimately, their good judgment of what matters most in incoming information streams, on social media and other online platforms, and to educate the public towards using online media parsimoniously and critically.

### 3- Connected products as a cause of depression, loss of sleep, altered brain function and myopia or early blindness in children

The use of connected devices and smartphones has increased rapidly in recent years, and this has brought about addiction. Studies investigating the relationship between smartphone use severity and sleep quality, depression, and anxiety in university students revealed that the Smartphone Addiction Scale scores of females were significantly higher than those of males. Depression, anxiety, and daytime dysfunction scores were higher in the 'high' smartphone user group compared with the 'low' smartphone user group. Positive correlations were found between the Smartphone Addiction Scale scores and depression levels, anxiety levels, and sleep quality scores **[8]**. In short, symptoms such as depression, anxiety, and poor sleep quality can be associated with smartphone overuse in student and other populations.

The recently observed increase in myopia and early blindness in children in Asia, Europe, and the United States has also been linked to overuse of connected devices. Social pressures prompting children to read for long hours on their computer and smartphone screens severely compromises time spent outdoors and significantly limits the child's exposure to sunlight, causing reduced levels of vitamin D in the body and, ultimately, long-term vitamine D deficiency. When chronic during a child's development, vitamine D deficiency may contribute to dysfunctions in the central nervous system. Moreover, long hours spent indoors on the computer reduce a child's far vision **[9]**. As a consequence, the eyes grow longer and become myopic. This disease progresses steadily once it has started, and ensues a high risk of severe early visual impairment, or early blindness. It was found that a significant increase in the time spent outdoors reduced the incidence of new myopia cases to half when children were sent to play outdoors without their phones/computers for at least 2 h daily. This positive effect of mandatory outdoor time is attributed to the light-induced retinal dopamine, which blocks the abnormal lengthening of the eyeball. In short, there is a clearly identified risk of early visual impairment or blindness in children as a result of long-hours of exposure to computer and smartphone screens. Regulatory measures and public awareness campaigns promoting a minimum of 2 hours outdoors activities all year long may help mitigate these risks. In addition, parents and educational institutions should be prompted towards limiting the time for online activity in children and teenagers to a minimum.

### 4- Internet of things: doomsday for individual data protection and privacy?

There are currently no measures that would allow us to predict how the next generation of technology devices, the so-called internet of things (Iot), will affect us psychologically. In order to work, Iot products have to see and/or listen to everything around them. If we buy enough of these devices, the Iot will be "listening in" on every aspect of our lives, from the kitchen to the car to the bedroom. While this will offer some convenience, the data collected by this technology will be used for things other than delivering the services the consumer paid for. Apart from a general "Big Brother is watching you" effect, which will affect some individuals psychologically more than others, this type of "eyes-and-ears-everywhere" technology could produce an unprecedented amount of misuse and abuse of consumer data by third parties. We may therefore expect negative psychological effects of a not yet clearly identified nature as a consequence of massive digitalisation and sensor data collection from everyone everywhere. These masses of data will be populating the cloud and create a genuine data jungle **[10]**, where an individual's private data become exposed use and misuse by any predator clever enough to hack them. The biggest risks here are ranking from misuse and abuse of personal data to straightforward identity theft, and it is not clear whether it will be possible at all to mitigate these risks within an appropriate regulatory framework.

## 5- *Virtual reality applications: how "real" does it get?*

There is growing evidence that virtual simulations of three-dimensional space can measurably affect the precision of human motor behaviours. Virtual reality simulations of 3D space (VR-3D) are used in interface technology for a variety of image-guided user tasks and are a choice target in the development of connected technology for '*the factory of the future'*. Tests on motor tasks with high-precision "to the single pixel" measures **[11]** have shown that VR-3D produces significantly less precise motor behaviours in comparison with high-fidelity 2D image guidance in operators not adapted to VR 3D. Moreover, evidence from multiple other sources suggests that over-adaptation to VR-3D may produce difficulties in readjusting motor responses to real-world space. This is explained by the fact that the human brain learns to adapt to 3D space over many years during ontogenetic development (especially during childhood, but we are talking about a lifelong adaptation process here) by finely tuned object-to-body calibration when reaching and grasping things**.** As a consequence, relative distances are not reliably assessed in VR-3D by the user/consumer **[12]**, and the adequate development strategy here would be to opt for high-quality 2D interfaces, especially when he precision of an image-guided motor behaviour is critical.

Another clearly identified problem with virtual reality applications is that of virtual *"doppelgangers"* in videogames and other applications. Doppelganger games were found to have measurable effects on an individual's cognitive ability, with memory loss and loss of control over his/her personal identity, in the game and potentially beyond. In the online virtual reality (VR) video game World of Warcraft **[13]** a player's avatar can be "mind-controlled" by other players. The target individual loses all control of his/her personal avatar, and must watch it being manipulated by another player, often with harmful intentions. Other VR video games allow for the building of characters that look like the user, and for algorithms to take over the behaviour of the user. Psychological studies on such autonomous "*doppelgangers"* have shown that this can lead to production of false memories and mental images; these can be manipulated by third parties without the consent or mental effort of the consumer **[14]**. Moreover, it is technically possible for one's digital "*doppelganger"* to exist and be manipulated long after the physical Self has died. This opens the door to a problem space, from ethically unacceptable misuse to abuse of players' or consumers' identities, where individuals no longer have control over what happens to their "*doppelgangers*" and/or what these latter may do to others. This urges to consider whether the development of technology using digital *"doppelgangers"* should be restricted, or even forbidden.

**Conclusions**

In a nutshell, AI and massive digitalisation incur certain risks in terms of reduced control over what we can do as individuals (to solve a specific problem, for example) and it may, at a deeper level, affect our perceptual and cognitive abilities and, thereby, daily behaviours as well as our sense of who we are as private human individuals. This can produce a general syndrome akin to "acquired helplessness" and may affect the capability of many individuals of coping with life at a larger scale. It has been demonstrated that a decreased belief in personal competence is sufficient to produce this syndrome in humans **[15]**. Also, massive digitalisation incurs the risk of creating an empathy-reduced society where, instead of spending more time interacting with our colleagues, friends, family and pets, we will be ultimately spending more and more time on our computers and connected devices instead. Apart from the addictive aspect of certain of these behaviours, it needs to be pointed out that living in empathy-reduced environments **[16]** is known to produce reduced levels of oxytocine, the *life and trust hormone* involved in the development of social behaviours that involve the feeling of trust **[17],** and increased levels of corticosterone, the *death and stress hormone* involved in cell death and premature ageing **[18]**. A chronic imbalance of the oxtocine/corticosterone regulation pathways in the human body can engender a variety of disorders in humans, ranging from dysfunctional behaviour caused by lack or loss of sleep to severe chronic fatigue, depression, and *burn-out* syndrome. A massive lack of human empathy in future societies may well not be the inevitable consequence of massive digitalisation, but there is a risk, and a trend in that direction is already beginning to show at the worldwide scale.